\documentclass[12pt]{article}
\usepackage[hypertex]{hyperref}
\usepackage{epsf}

\def\Omit#1{}

\newcommand{\im}{{\rm Im}}

\newcommand{\beq}{\begin{equation}}
\newcommand{\eeq}{\end{equation}}
\newcommand{\ba}{\begin{array}}
\newcommand{\ea}{\end{array}}
\newcommand{\beqa}{\begin{eqnarray}}
\newcommand{\eeqa}{\end{eqnarray}}

\newcommand{\cF}{{\cal F}}
\newcommand{\cG}{{\cal G}}

\newcommand{\br}{{\cal B}}

\newcommand{\no}{\nonumber}
\newcommand{\lsim}{\stackrel{<}{_\sim}}

\def\npb#1#2#3{    {\it Nucl. Phys. }{\bf B #1} (#2) #3}
\def\plb#1#2#3{    {\it Phys. Lett. }{\bf B #1} (#2) #3}

\def\prd#1#2#3{    {\it Phys. Rev. }{\bf D #1} (#2) #3}

\def\prl#1#2#3{    {\it Phys. Rev. Lett. }{\bf #1} (#2) #3}

\def\zpc#1#2#3{    {\it Z. Phys. }{\bf C #1} (#2) #3}

\def\ibid#1#2#3{   {\it ibid. }{\bf #1} (#2) #3}
\def\jhep#1#2#3{   {\it JHEP  }{\bf #1} (#2) #3}

\begin{document}

\thispagestyle{empty}
\begin{flushright}
SLAC-PUB-10254 \\
hep-ph/0311353
\end{flushright}
\vskip 1.0 true cm

\begin{center}
{\Large\bf Lepton-flavor mixing and $K \to \pi \nu\bar\nu$
           decays }\\ [35 pt]
{Yuval Grossman,${}^{a,b,c}$ Gino Isidori,${}^{d}$ and 
Hitoshi Murayama${}^{e}$\footnote{On leave of absence from Department
of Physics, University of California, Berkeley, CA 94720, USA}} 
 \\ [15 pt]

{\sl ${}^a$Technion--Israel Institute of Technology, Technion City, 32000
Haifa, Israel \footnote{Permanent address}}
\vspace{5pt}\\
{\sl ${}^b$Stanford Linear Accelerator Center, 
  Stanford University, Stanford, CA 94309, USA}
\vspace{5pt}\\
{\sl ${}^c$ Santa Cruz Institute for Particle Physics, 
University of California, Santa Cruz, CA 95064, USA}
\vspace{5pt}\\
{\sl ${}^d$ INFN, Laboratori Nazionali di Frascati,
 00044 Frascati, Italy} \\ [5pt]
{\sl ${}^e$ School of Natural Sciences, Institute for Advanced Study,
Princeton, NJ 08540, USA} \\ [30pt]
{\bf Abstract}
\end{center}
\noindent
The impact of possible sources of lepton-flavor mixing on $K \to \pi
\nu\bar\nu$ decays is analysed. At the one-loop level lepton-flavor mixing
originated from non-diagonal lepton mass matrices cannot generate a
CP-conserving $K_L \to \pi^0\nu\bar\nu$ amplitude.  The rates of these
modes are sensitive to leptonic flavor violation when there are
at least two different leptonic mixing matrices. New interactions that
violate both quark and lepton universalities could enhance the
CP-conserving component of $\Gamma(K_L \to \pi^0\nu\bar\nu)$ and have
a substantial impact. Explicit examples of these effects in the
context of supersymmetric models, with and without $R$-parity
conservation, are discussed.

\newpage

\section{Introduction}
Within the Standard Model (SM), the Flavor-Changing-Neutral-Current
(FCNC) decays $K\to\pi\nu\bar{\nu}$ are among the cleanest observables
to determine the mixing of the top quark with the light
generations. In particular, the $K_L \to\pi^0 \nu\bar{\nu}$ rate is
completely dominated by a CP-violating (CPV) amplitude and could be
used to determine with high precision the Jarlskog's invariant
\cite{Litt,BB1}. The situation could be very different beyond the SM:
similarly to all FCNC transitions, $K\to\pi\nu\bar{\nu}$ decays are
highly sensitive to new sources of quark-flavor mixing. However, a
peculiar aspect of these decays is their potential sensitivity also to
flavor mixing in the leptonic sector. The most remarkable consequence
of this fact is that the transition $K_L \to\pi^0 \nu_i \bar{\nu}_j$,
with $i\not=j$, does not need to be dominated by a CPV
amplitude~\cite{GN}.

Recent results from neutrino physics indicate that the quark and
lepton sectors have a rather different flavor structure. In
particular, we now know that large mixing angles do appear in the
lepton sector. Due to the smallness of neutrino masses, these large
mixing angles have no impact on $K \to \pi \nu\bar\nu$ rates in
minimal models, where only neutrino mass terms are introduced
\cite{Perez}. However, this conclusion is not necessarily true in more
general scenarios, such as supersymmetric models, with possible large
mixing angles also in the slepton sector.

In this letter we present a general analysis of the impact of
lepton-flavor mixing on $K \to \pi \nu\bar\nu$ decays. As we shall
show, if left-handed neutrinos are the only light fields and
lepton-flavor mixing is confined only to mass matrices, lepton-flavor
violation cannot be the dominant effect on the $K\to\pi\nu\bar{\nu}$
rates. In particular, it cannot induce a CPC $K_L \to\pi^0 \nu \bar{\nu}$ amplitude.
This conclusion is independent of the type of mass matrices
involved (e.g., neutrinos, sneutrinos, leptons or sleptons).
However, if more than one mass matrix is involved,
the effect of lepton-flavor mixing is not necessarily negligible. 
We demonstrate it in the Minimal Supersymmetric SM (MSSM), where the 
charged-slepton-neutrino and the sneutrino-neutrino mixing matrices are in
general different. In order to induce a non-negligible 
CPC $K_L \to\pi^0 \nu \bar{\nu}$ transition, lepton-flavor mixing 
in mass matrices is not sufficient and we need a 
new interaction that violates both quark and 
lepton universality. 
We illustrate this effect with two examples of non-universal
interactions: the lepton-quark-squark coupling in the framework of the
$R$-parity violating MSSM and the Yukawa interaction in the $R$-parity
conserving MSSM.

\section{General properties of  $K\to\pi\nu\bar{\nu}$ amplitudes}
The SM contributions to $K\to\pi\nu\bar{\nu}$ amplitudes
are described by the following effective Hamiltonian \cite{BB2}
\beq
{\cal H}^{\rm SM}_{\rm eff} = \frac{4 G_F}{\sqrt 2} \frac{\alpha}{2\pi
\sin^2\Theta_W} \sum_{\ell=e,\mu,\tau} \left[ \lambda_c X^\ell_{NL} +
\lambda_t X(x_t) \right]~ \bar s_L \gamma^\mu d_L \times \bar\nu^\ell_L
\gamma_\mu \nu^\ell_L +{\rm h.c.}~,
\label{eq:Heff}
\eeq
where $x_t=m_t^2/M_W^2$, $\lambda_q = V^*_{qs}V_{qd}$ and $V_{ij}$
denote CKM matrix elements. The coefficients $X^\ell_{NL}$ and
$X(x_t)$, encoding top- and charm-quark loop contributions, are known
at the NLO accuracy in QCD \cite{BB2,MU} leading to a very precise
prediction of the decay rates. Note that the dependence on the lepton
flavor that enter via $X^\ell_{NL}$ is very small, and we neglect it
in the following. The neutrino pair produced by ${\cal H}^{\rm
SM}_{\rm eff}$ is a CP eigenstate with positive eigenvalue. This is
the reason why the leading SM contribution to $K_L \to \pi^0 \nu_i
\bar
\nu_i$ is due to CP violation~\cite{Litt}.  Within the SM,
CP-Conserving (CPC) contributions to $K_L \to \pi^0 \nu_i \bar \nu_i$
are generated only by local operators of dimension $d \geq 8$ or by
long-distance effects: these contributions do not exceed the $10^{-4}$
level in the total rate, compared to the dominant CP-violating term~\cite{BI}. 

The situation could be very different beyond the SM, where new
dimension-six operators could contribute to $K \to \pi \nu \bar \nu$
amplitudes. In principle, beyond the SM one should also take into
account other $K \to \pi + X_{\rm invisible}$ transitions, which could
lead to similar experimental signatures.  In order to classify the
relevant operators, it is necessary to specify which are the light
invisible degrees of freedom of the theory, and what are their
interactions. For our purpose, we can distinguish three main
scenarios:
\begin{enumerate}
\item{} {\em The only light invisibles are the three species of left-handed neutrinos}.\\
In this case the only relevant dimension-six operators are:
\beq
O_{sd}^{ij} = \bar s \gamma_\mu d \times \bar \nu^i_L \gamma^\mu \nu_L^j~.
\label{eq:op}
\eeq
For $i\not =j$ these operators create a neutrino pair 
which is not a CP eigenstate. In principle, one can also write operators 
of the type $(\bar s \Gamma d) \times \nu_L^C \Gamma \nu_L$,
which break both lepton-number and $SU(2)_L$-invariance. As expected by this 
highly-breaking structure, and as explicitly shown in  Ref.~\cite{Perez},
the effect of these additional operators is completely negligible. 

\item{} {\em Right-handed neutrinos are also light, but they are
    sterile.}\\
In this case we need to consider also scalar and tensor
dimension-six operators of the type 
$(\bar s \Gamma d) \times  \bar{\nu}_{R(L)} \Gamma \nu_{L(R)}$;
however, if right-handed neutrinos are sterile the coupling of 
these operators is negligible. An explicit realization of this 
scenarios occurs in all the models where the right-handed neutrinos
interact with the SM fields only through their 
(tiny) Dirac mass terms \cite{Perez}.

\item{} {\em Right-handed neutrinos are light and not sterile.}\\
If the right-handed neutrino fields are not sterile, the coupling 
of the scalar and tensor operators mentioned above (case 2) 
is not necessarily suppressed and these operators 
could compete with the leading
left-handed terms in (\ref{eq:op}). This occurs for instance in 
LR symmetric models, where the right-handed neutrino 
fields couple to quarks via new gauge interactions \cite{LR}.
In this framework lepton-flavor mixing could have a 
non-negligible impact on $K \to \pi \nu \bar \nu$ rates. 
The scalar and tensor operators have a different CP 
structure with respect to the SM operator and they 
induce a CPC contribution to $K_L \to \pi^0 \nu \bar \nu$ 
in absence of lepton-flavor mixing \cite{LR}.
\end{enumerate}

An important difference of the last two cases with respect to the
first one is the fact that scalar and tensor operators would also lead
to a different pion-energy spectrum. Thus, the first case is in
principle distinguishable from the last two by means of
experimental data.  This conclusion can be generalized to most of the
other $K \to \pi + X_{\rm invisible}$ transitions, where $X_{\rm
invisible}$ include other degrees of freedom in addition to the
neutrinos.\footnote{For example, there is a possible decay $K
\rightarrow \pi f$ where $f$ is a ``familon,'' a Nambu--Goldstone
boson of the spontaneously broken horizontal symmetry.  However,
this process can be discriminated experimentally because of the
two-body kinematics.}

In the following, we shall analyze in more detail the effect of
lepton-flavor mixing in the first case above, when only the operators
(\ref{eq:op}) are relevant, and then the pion-energy spectrum is
identical to the SM case.

\section{Lepton-flavor mixing in mass matrices}

Since the $\nu^j \bar \nu^i$ final state is not a CP eigenstate, 
the condition for a non-vanishing $K_L \to \pi^0 \nu \bar
\nu$ rate seems to be the breaking of CP or lepton-flavor
symmetries. As we explain below, the condition turns out to be
stronger: we need either CP violation in the quark sector or a 
new effective interaction that violates both quark and 
lepton universality. 

If the breaking of flavor universality can be confined only 
to appropriate mass matrices, both in the quark and in the lepton sectors, 
and the two sectors are connected by flavor-universal interactions,
quark- and lepton-flavor mixing terms in $K \to \pi \nu\bar\nu$ 
amplitudes assume a factorizable structure. In this case we 
can always rotate the neutrino eigenstates to diagonalize the
lepton final state, without any impact on the quark strucutre. 
As a result, the inclusive sum over neutrino flavors can be 
transformed into a sum over CP eigenstates. It is then clear that 
the $K_L \to \pi^0 \nu \bar\nu $ transition vanishes 
in absence of CP violation in the quark sector.  

\begin{figure}[t]
    \begin{center}
       \setlength{\unitlength}{1truecm}
       \begin{picture}(10.0, 4.0)
       \epsfxsize 10.  true cm
       \epsffile{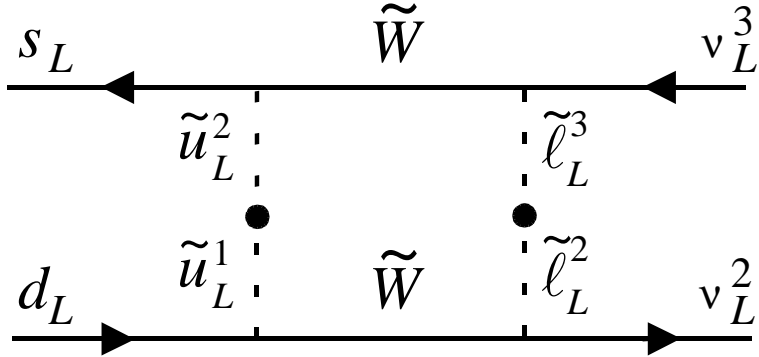}
       \end{picture}
    \end{center}
    \vskip -0.5 cm
    \caption{Wino-Wino box diagram.}
    \protect\label{fig:boxww}
\end{figure}

We note the following two points:
\begin{enumerate}
\item Even with the factorizable structure, the (lepton) mass matrices
  may have impact on $K \to \pi \nu\bar\nu$ rates.  The eigenvalues of
  the mass matrices are certainly relevant and, if more than one
  non-trivial mass matrix is involved, also their relative rotation
  angles can play a significant role.
\item The factorization structure is expected to be broken by
  higher-order loop effects. Then, the flavor breaking in the mass
  terms could induce a breaking of universality also in effective
  interaction vertices. Since this is a higher-order effect, it is
  likely to be highly suppressed.
\end{enumerate}

\begin{figure}[t]
    \begin{center}
       \setlength{\unitlength}{1truecm}
       \begin{picture}(10.0, 4.0)
       \epsfxsize 10.  true cm
       \epsffile{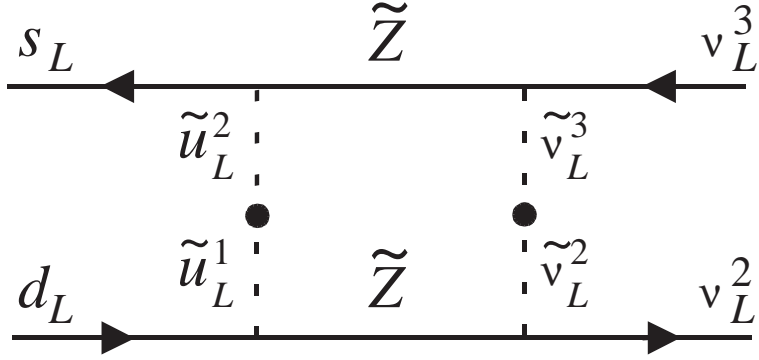}
       \end{picture}
    \end{center}
    \vskip -0.5 cm
    \caption{Zino-Zino box diagram.}
    \protect\label{fig:boxzz}
\end{figure}

To illustrate the above argument, we discuss a specific example of a
factorizable structure: the one originated from the $\widetilde W$-box
diagrams in Fig.~\ref{fig:boxww}.
Using the fact that the weak interaction is universal, we can write the
decay amplitude in the basis where squark and slepton mass matrices are
diagonal as
\beq
A(K^0 \to \pi \nu_i \bar \nu_j) = \frac{1}{\sqrt{2}}
\sum_{q,\ell} {\hat V}_{sq} {\hat V}_{dq}^* {\hat U}_{i\ell}
{\hat U}_{j\ell}^* f(m_{\tilde q},m_{\tilde \ell})~.
\label{eq:UV}
\eeq
Here $f(m_{\tilde q},m_{\tilde \ell})$ is the loop function, which
depends on squark and slepton masses. 
${\hat V}$, [${\hat U}$] is
a unitary matrix describing the rotation from the electroweak
(interaction) eigenstates to the mass eigenstates in the $\widetilde
W \widetilde u_i d_j$ [$\widetilde
W \widetilde \ell_i \nu_j$] interaction. 
Working in the basis where $CP|K^0\rangle=|\bar K^0\rangle$ we get
\beq
A(\bar K^0 \to \pi \nu_i \bar \nu_j) = \frac{1}{\sqrt{2}}
\sum_{q,\ell} {\hat V}_{sq}^* {\hat V}_{dq} {\hat U}_{i\ell}
{\hat U}_{j\ell}^* f(m_{\tilde q},m_{\tilde \ell})~.
\label{eq:UVbar}
\eeq
Introducing the diagonal matrix $\cF_q$,
defined by $(\cF_q)_{ii}=f(m_q,m_i)$, the $K_L \to \pi^0 \nu_i \bar
\nu_j$ amplitude can be written as
\beq \label{Famp}
A(K_L \to \pi \nu_i \bar \nu_j) = i
\sum_{q} \im \left({\hat V}_{sq} {\hat V}_{dq}^*\right)
\left[ {\hat U} \cF_q {\hat U}^\dagger \right]_{ij}~.
\eeq
We see that the amplitude vanishes if there is no CP-violation in the quark
sector. 

Assuming that the full amplitude is given by eq. (\ref{Famp}) and
ignoring phase-space effects due to non-vanishing neutrino masses, the
total rate obtained by summing over neutrino flavors is given by
\beqa
\Gamma( K_L \to \pi^0 \nu \bar \nu) &\propto&
\sum_{ij} \left|A(K_L \to \pi \nu_i \bar \nu_j)\right|^2
 \\ &=&
\sum_{q,k} \im \left({\hat V}_{sq} {\hat V}_{dq}^*\right)
             \im \left({\hat V}_{sk} {\hat V}_{dk}^*\right)~ {\rm tr}
             \left[ {\hat U} \cF_q {\hat U}^\dagger {\hat U} \cF_k
             {\hat U}^\dagger \right] \nonumber \\ &=&
\sum_{q,k} \im \left({\hat V}_{sq} {\hat V}_{dq}^*\right)
             \im \left({\hat V}_{sk} {\hat V}_{dk}^*\right)~ {\rm tr}
             \left[\cF_q \cF_k \right] \,.
\nonumber 
\eeqa
We see that the lepton-flavor mixing matrix ${\hat U}$ disappears from
the trace over lepton indices.  This is a result of the fact that we
sum over all the final-state neutrino flavors.  On the other hand, the
eigenvalues of the slepton mass matrix enter in the determination of
${\rm tr}[\cF_q \cF_k]$.

Similar arguments hold also for the SM with massive neutrinos
\cite{Perez}. In that case, as well as in our more general case, the
$K_L$ decay amplitude arises only due to CP violation in the
quark (or squark) sector.

Now we consider a case where we have two different amplitudes with
different flavor mixing. For example, we add the $\widetilde Z$-box
diagrams of Fig.~\ref{fig:boxzz}. Similarly to the wino diagram the
amplitude is given by
\beq \label{Famp-z}
A(K_L \to \pi \nu_i \bar \nu_j) = i
\sum_{q} \im \left({\hat V'}_{sq} {\hat V}_{dq}^{\prime*}\right)
\left[ {\hat U'} \cG_q {\hat U}^{\prime\dagger} \right]_{ij}~.
\eeq
where $\cG_q$ is defined similar to $\cF_q$ and
the primed matrices are the ones that rotate the neutral
interaction. In general, $V\ne V'$ and $U\ne U'$. Adding the two
amplitudes and neglecting the SM contribution we get 
\beq
\Gamma(K_L \to \pi^0 \nu \bar \nu) \propto
\sum_{q,k}\left\{ a^q_F a^k_F {\rm tr}\left[\cF_q \cF_k \right] + 
a^q_G a^k_G {\rm tr}\left[\cG_q \cG_k \right] + 
2 \, a^q_F a^k_G {\rm tr}\left[W^\dagger \cF_q W \cG_k \right] \right\}, 
\eeq
where
\beq
a_F^q= \im \left({\hat V}_{sq} {\hat V}_{dq}^*\right), \qquad
a_G^q= \im \left({\hat V'}_{sq} {\hat V}_{dq}^{\prime*}\right)~,
\eeq
and 
\beq
W\equiv \hat U^\dagger \hat U'~.
\eeq 
We see that the product of the mixing matrix enter in the
interference term.

We note the following points:
\begin{enumerate}
\item
When $\cF_q$ or $\cG_q$ are proportional to the unit matrix there is no
sensitivity to the mixing matrix $W$. This is the case when the charged
sleptons or sneutrinos are degenerate. More generally, we conclude that
the effect is suppressed by the amount of degeneracy in the slepton sector.
\item
The effect of the leptonic mixing cannot be very large. Since $W$ is
unitary, we learn that it is at most an $O(1)$ effect. Yet, the effect
can be large enough to be detectable. 
\end{enumerate}

\vfill

\section{Flavor non-universal interactions}

\subsection{$R$-parity violating SUSY}

A typical example of interaction that violates both quark and lepton
universality is a family non-universal leptoquark (LQ).  In $R$-parity
violating supersymmetric models, the squarks, which couples to quark
and leptons via the $R$-parity breaking $LQ\bar d$ term, provides an
explicit example of this scenario. In this context, the CPC $K_L \to
\pi^0 \nu_i \bar\nu_j$ transition mediated by operators of the type
(\ref{eq:op}) is generated already at tree level. To illustrate the
general conditions under which the CPC rate can be large, we shall
discuss the LQ example in more detail.

Consider the following interaction term
\beq
\lambda_{\ell q}\, \bar q^{\,c}_L \ell_L S~,
\label{eq:intLQ}
\eeq
where $S$ is a scalar LQ and the other notations are clear.  This
leads to the following effective dimension-six Hamiltonian \cite{gln}
\beqa
{\cal H}^{\rm LQ}_{\rm  eff} &=& \frac{1}{M_S^2}
\Big\lbrace  \;  \bar s \gamma_\mu d ~
\left[ \lambda_{i s}\lambda^*_{j d} ~ \bar \nu^i_L \gamma^\mu \nu_L^j~ +
   \lambda_{j s}\lambda^*_{i d} \bar \nu^j_L \gamma^\mu \nu_L^i \right] \no\\
      &&\qquad  +
 \bar d \gamma_\mu s ~
\left[ \lambda_{i d}\lambda^*_{j s} ~ \bar \nu^i_L \gamma^\mu \nu_L^j~ +
   \lambda_{j d}\lambda^*_{i s} \bar \nu^j_L \gamma^\mu \nu_L^i
      \right]  \Big\rbrace~.
\label{eq:Heff-LQ}
\eeqa
We then obtain
\beq
A(K_L \to \pi \nu_i \bar \nu_j) \propto
(\lambda_{i s}\lambda^*_{j d} - \lambda_{i d}\lambda^*_{j s}).
\eeq

We note the following points:
\begin{enumerate}
\item
In the general case there is no lepton and quark factorization. Then
the $K_L$ amplitude does not vanish, and, for $i \ne j$, contains both
CPV and CPC terms.
\item
In a specific scenario where the LQ coupling is universal with respect
to the lepton flavor, namely $\lambda_{i q}=\lambda_{j q}$ for each
$q$, the amplitude is proportional to $\im(\lambda_{i s}\lambda^*_{i
d})$. In this case the amplitude is purely CP violating where,
similarly to the SM case, the CP violation originates from the
quark sector.
\item
If the LQ coupling is universal with respect to the quark flavor,
namely $\lambda_{i s}=\lambda_{i d}$ for each $i$, the amplitude
vanishes. This is expected since quark mixing is necessary for any
FCNC process.
\end{enumerate}

In the case of quarks and leptons of the first two generations, the
interaction term in (\ref{eq:intLQ}) is severely constrained by $\pi$
and $K$ semileptonic decays. Nonetheless, the strongest bound on
off-diagonal combinations like $\lambda_{2 s}^*\lambda_{3 d}$ come
from $B(K^+\to\pi^+\nu\bar\nu)$ \cite{GN,Davidson}. Therefore, tuning
appropriately these parameters one can generate a huge CPC conserving
transition of the type $K_L \to \pi^0 \nu_3 \bar\nu_2 + \pi^0\nu_2
\bar\nu_3$. As mentioned before, this occurs only when $\lambda_{2
  s}\lambda^*_{3 d} \ne \lambda_{2 d}\lambda^*_{3 s}$. In that case
the final state is not a CP eigenstate and thus the decay is a
combination of CPC and CPV transitions.

As shown in \cite{GN}, also in this scenario
the $K_L$ width  (summed over neutrino flavors)
cannot exceed in magnitude the $K^+$ one.
In view of the recent BNL-E787 result on the charged mode
\cite{E787}, this model-independent relation implies
$\br(K_L\to\pi^0\nu\bar{\nu}) <  1.7 \times 10^{-9}~
(90\%~{\rm C.L.})$.

\subsection{$R$-parity conserving SUSY}
In less exotic scenarios, like the SM with massive neutrinos or the
MSSM with $R$-parity conservation, the only interaction that violates
quark and lepton universality is the Yukawa interaction. Therefore,
within these models the CPC contributions to $K_L \to \pi^0 \nu_i
\bar\nu_j$ are necessarily suppressed by Yukawa couplings. In the SM
these terms are absolutely negligible \cite{Perez}. The situation,
however, is less obvious in the MSSM. There, the Yukawa couplings of
the lepton (for $\tan\beta\gg1$) and both the slepton- and
squark-flavor mixing angles can be large.

\begin{figure}[t]
    \begin{center}
       \setlength{\unitlength}{1truecm}
       \begin{picture}(10.0, 4.0)
       \epsfxsize 10.  true cm
       \epsffile{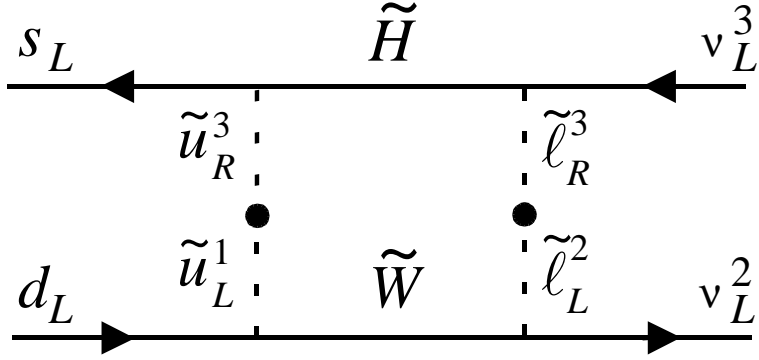}
       \end{picture}
    \vskip -0.5 cm
    \end{center}
    \caption{Wino-Higgsino box diagram}
    \protect\label{fig:boxhw}
\end{figure}

The potentially largest CPC contribution is generated from the
non-universal interaction in Fig.~\ref{fig:boxhw}. Contrary to the
case of Fig.~\ref{fig:boxww}, here the exchange $s\leftrightarrow d$
cannot be simply re-absorbed into the phase of the quark-mixing term.
{}From the point of view of the low-energy effective Hamiltonian, this
diagram is equivalent to a LQ exchange with
\beq
\lambda_{2s}\lambda^*_{3d}
\propto y_t y_\tau V_{ts}^* (\delta^{U*}_{LR})_{1 3} (\delta^L_{LR})_{2
3}.
\eeq
Similar contribution arises for the $(\nu_3\bar\nu_3)$ final state, but
then the amplitude is proportional to
$\im(V_{ts}^*(\delta^{U*}_{LR})_{1 3})$ and the effect is purely CPV.

Considering only the flavor violating contribution and
using the results of Refs. \cite{Kpnn_SUSY,BRS} we obtain
\beqa
\frac{ \Gamma(K_L \to \pi^0 \nu_3 \bar\nu_2 + \pi^0\nu_2 \bar\nu_3)_{\rm MSSM} }{
\Gamma(K_L \to \pi^0 \nu\nu)_{\rm SM}} &=& \frac{1}{3}
\left( \frac{m_\tau m_t \tan\beta}{8 M^2_{\widetilde W}} \right)^2
\left| \frac{ (\delta^U_{LR})_{1 3} (\delta^L_{LR})_{23} }{\im (V_{td})_{\rm SM} } \right|^2
\left| \frac{F_{\rm loop}(x_{ij}) }{X_t(x_t)} \right|^2~
\no\\
&\lsim&
\left( \frac{\tan\beta}{50} \right)^2 \left|
(\delta^U_{LR})_{1 3} (\delta^L_{LR})_{23} \right|^2~.
\label{eq:CPC1}
\eeqa
As usual $(\delta^A_{LR})_{i j} = ({\tilde M}^2_A)_{i_L j_R}/ ({\tilde
M}^2_A)_{i_L i_L}$ denote off-diagonal entries of squark and lepton
mass matrices. The dimensionless loop function $F_{\rm loop}(x_{ij})$,
which depends on the ratio of sparticle masses ($x_{ij}=m_i^2/m_j^2$) is
very small:
\beq
F_{\rm loop}( x_{ij} )  = x_{q_L \chi_1} x_{\ell_L \chi_1}
k( x_{q_L \chi_1},  x_{q_R \chi_1}, x_{\ell_L \chi_1},
x_{\ell_R \chi_1}, x_{\chi_2 \chi_1} )
~  \longrightarrow ~\frac{1}{30} \quad (\mbox{for}~x_{ij}=1)~,
\eeq
with $k$ defined as in \cite{BRS}.  This confirms the observation of
Ref.~\cite{Kpnn_SUSY} that SUSY box-diagram contributions to $K \to
\pi \nu \bar\nu $ are suppressed.
The upper figure ($F_{\rm loop} \approx 0.05$) is obtained with a
large splitting between left-handed and right-handed sfermions.

Given the bounds on the left-right mass insertions of squarks
\cite{Kpnn_SUSY,BRS} and leptons \cite{LRbounds}, we conclude that
the ratio in Eq.~(\ref{eq:CPC1}) cannot exceed the $10^{-2}$ level.
If this bound were saturated, this CPC contribution would be much
larger that the SM one, but of course would still be negligible
compared to the SM CP-violating rate (and thus undetectable).

\section{Conclusions}
$K\to\pi\nu\bar{\nu}$ decays are certainly one of the cleanest
windows to the short-distance mechanism of quark-flavor mixing. The
result of the BNL--E787 Collaboration \cite{E787}, although still
affected by a large experimental error, already shows the great
potential of these modes in constraining flavor physics within and
beyond the SM \cite{DI}.

Beside the obvious sensitivity to quark-flavor mixing,
$K\to\pi\nu\bar{\nu}$ decays are in principle affected also by mixing
of lepton flavors \cite{GN}.  In this letter we have investigated
under which conditions the leptonic mixing can play a significant role
in these modes. First we studied the case where the sources of quark-
and lepton-flavor mixing can be factorized.  In particular, we
concentrate on cases where the source of flavor-symmetry breaking is
confined to mass matrices, since then this factorization is almost
complete.  We found that the sum over neutrino flavors (implicitly
understood in $K\to\pi\nu\bar{\nu}$ rates) wash out any individual
effect due to lepton-flavor mixing. Only in cases where there are two
different lepton-flavor mixing matrices, there is an effect which
depends on the product of the two mixing matrices. Then we studied
interactions that violates at the same time quark and lepton
universality. In that case individual
leptonic flavor violation can be important as they
induce CPC contribution to the rate.

In models like the SM or the $R$-parity conserving MSSM, but also in
models with large extra dimensions with a protective flavor symmetry
(see e.g. Ref.~\cite{AHH}), only the Yukawa interaction violates at
the same time quark and lepton universality. In these models
CPC lepton-flavor mixing effects in $K\to\pi\nu\bar{\nu}$ decays are
therefore suppressed by Yukawa couplings. As we have explicitly shown,
even in a very favorable case, such as the MSSM with generic flavor
couplings and large $\tan\beta$, these types of CPC lepton-flavor mixing
effects are negligible. In more exotic
scenarios, such as $R$-parity violating supersymmetric models,
lepton-flavor mixing could generate significant effects in
$K\to\pi\nu\bar{\nu}$ decays, in particular, a sizable $K_L
\to\pi^0\nu\bar{\nu}$ CP-conserving rate.

\newpage

\section*{Acknowledgments}

We thank the organizers of the Workshop On CKM Unitarity Triangle at
CERN, 13-16 Feb 2002, Geneva, Switzerland, where this work has
started.  We thank Yossi Nir and Gilad Perez for useful discussions.
The work of YG is supported in part by the Department of Energy,
contract DE-AC03-76SF00515 and by the Department of Energy under grant
no.~DE-FG03-92ER40689.  GI acknowledges the hospitality of the Theory
Group at SLAC, where part of this work has been done, and a partial
support from the EC IHP-RTN program, contract no.~HPRN-CT-2002-00311
(EURIDICE).  The work of HM was supported by Institute for Advanced
Study, funds for Natural Sciences.  HM was also supported in part by
the DOE Contract DE-AC03-76SF00098 and in part by the NSF grant
PHY-0098840.

\end{document}